\documentclass[aps,pra,11pt,letterpaper,reprint,notitlepage,superscriptaddress,floatfix]{revtex4-2}
\usepackage{amsmath,amssymb,amsfonts}
\usepackage[bookmarks=false]{hyperref}
\usepackage{physics}
\usepackage{lmodern}
\usepackage{keyval}

\usepackage{graphicx}

\usepackage[capitalise]{cleveref}

\usepackage[range-units=single,separate-uncertainty]{siunitx}

\renewcommand{\d}{\dd}
\newcommand{\avg}[1]{\left\langle {#1}\right\rangle}

\renewcommand{\t}[1]{\mathrm{#1}}

\newcommand{\LigoMIT}{LIGO Laboratory, Massachusetts Institute of Technology, Cambridge, MA 02139}
\newcommand{\UChic}{The James Franck Institute and Department of Physics,
University of Chicago, Chicago, Illinois 60637}
\newcommand{\Stanf}{Department of Physics, Stanford University, Stanford, California 94305}
\newcommand{\MechMIT}{Department of Mechanical Engineering, Massachusetts Institute of Technology, Cambridge, MA 02139}

\newcommand{\PhysVienna}{Faculty of Physics, University of Vienna, Boltzmanngasse 5, 1090 Vienna, Austria}
\newcommand{\IQOQI}{Institute for Quantum Optics and Quantum Information (IQOQI Vienna) of the Austrian Academy of Sciences, Boltzmanngasse 3, A-1090, Vienna, Austria}

\begin{document}

\title{Limits and prospects for long-baseline optical fiber interferometry}

\author{Christopher Hilweg}
\email{christopher.hilweg@univie.ac.at}
\affiliation{\PhysVienna}
\affiliation{\IQOQI}
\author{Danial Shadmany}
\thanks{The first two authors contributed equally to this work.}
\affiliation{\UChic}
\affiliation{\Stanf}
\author{Philip Walther}
\affiliation{\PhysVienna}
\author{Nergis Mavalvala}
\affiliation{\LigoMIT}
\author{Vivishek Sudhir}
\email{vivishek@mit.edu}
\affiliation{\LigoMIT}
\affiliation{\MechMIT}

\begin{abstract}
Today's most precise optical instruments --- gravitational-wave interferometers and optical atomic clocks --- rely on long storage times for photons to realize their exquisite sensitivity. 
Optical fiber technology is the most widely deployed platform for realizing long-distance optical propagation. Yet, their application to precision optical measurements is sparse. We review the state-of-the-art in the noise performance of conventional (solid-core) optical fibers from the perspective of precision optical measurements and quantum technology that rely on precise transfer of information over long distances. In doing so, we highlight the limitations of this platform and point to the opportunities that structured fiber technology offers to overcome some of these limitations.
\end{abstract}
\maketitle


\section{Introduction}\label{sec:Intro}

An ancient material, sand, underlies much of the technological prowess of the modern world. In particular, the information age that we live in is enabled by the fact that ultra-pure fused silica can sustain ultra-low-loss optical propagation \cite{Kao66,Hecht}. A century-long pursuit in ceramic science and  glass technology \cite{Doug66,KurkMat89,Mac90,DraBal18}, together with theoretical understanding of optical loss mechanisms in these materials \cite{Kao66,Olsh79}, has resulted in optical fibers  limited only by intrinsic losses as low as $\approx \SI{0.14}{dB/km}$ (at a wavelength of \SI{1.55}{\micro m}) \cite{TamHas17}. Together with the laser, low-loss optical fibers enable the delivery of highly coherent radiation across large distances.

Indeed optical loss is a convenient figure of merit to gauge the maturity of an optical technology and a relevant one for its application in both classical and quantum technologies. Today's most precise measurements --- incidentally, of space \cite{Tse19,Buik20} and time \cite{ZheKolk21,BotYe21}, at a precision around
1 part in $10^{20}$ --- are carried out using ultra-low-loss optical interferometers. Here the requirement for low optical loss arises from the need to evade sensitivity limitations arising from quantum fluctuations in the optical field \cite{BragKip00}. Optical quantum networks are another example, where the effect of loss is two-fold: loss of photons that carry quantum information is a mechanism of direct decoherence, whereas losses in the classical communication channel required to complete a quantum communication protocol limit the fidelity with which the latter can be performed \cite{BartSpek07}. In both cases, reduction of losses and extraneous noise are central to the scientific goals they serve.

In order to appreciate how the humble optical fiber compares against these state-of-the-art scientific instruments, consider the following. In terms of the ability to store photons --- a direct measure of the sensitivity of an interferometer \cite{BragKip00} --- Fabry-Perot cavity interferometers, such as the ones employed in  Advanced LIGO or as references in optical atomic clocks, realize losses as low as $\approx \SI{10}{ppm/km}$ \cite{brooks2021point}, which is apparently three orders of magnitude lower than the state-of-the-art loss
of contemporary solid-core optical fibers \cite{Tamura2017,Tamura2018}. However that performance is achieved by elongating the cavity, which ``dilutes'' the fraction of time the photon interacts with the absorptive surface of the cavity mirrors. A direct comparison of optical loss for photons living entirely inside a transparent medium paints a different picture: losses in the few-micron-thick mirror coatings of interferometer cavities can be as large as several tens of dB/km, orders of magnitude worse than the pristine fused silica core of an optical fiber. 

\begin{figure}[t!]
    \centering
    \includegraphics[width=\columnwidth]{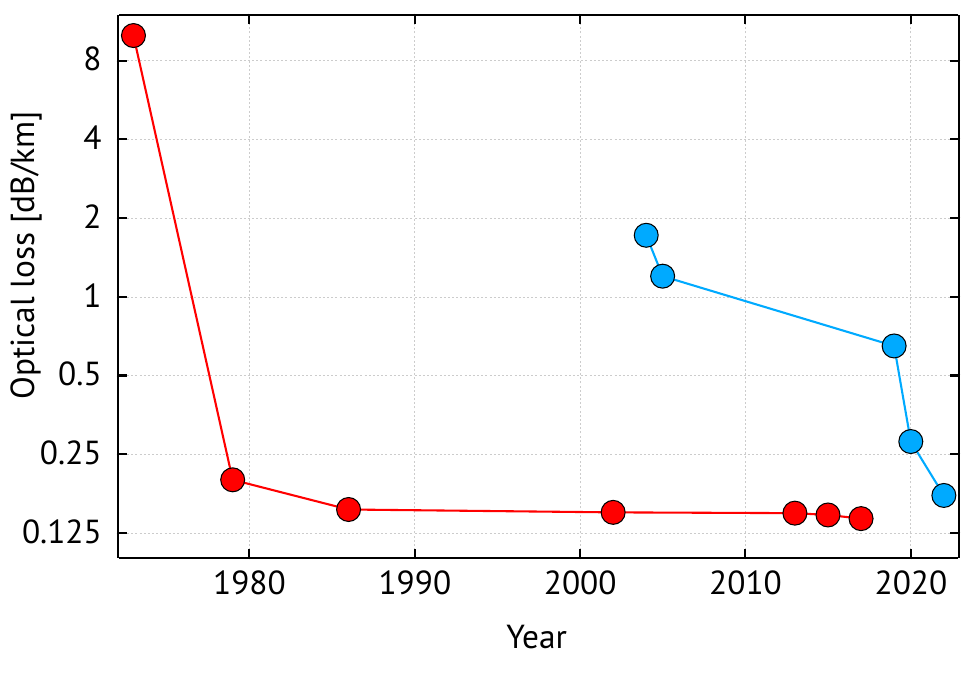}
    \caption{\label{Fig:Losses}
        Evolution of fiber attenuation for solid- (purple) and hollow-core fibers (orange) for wavelengths in the telecom C-band (lines to guide the eye). Solid-core fibers improved from about 10 dB/km (1973) \cite{Keck1973}, 0.2 dB/km (1979) \cite{Miya1979}, 0.154 dB/km (1986) \cite{Kanamori1986}, 0.15 dB/km (2002) \cite{Nagayama2002}, 0.149 dB/km (2013) \cite{Hirano2013}, 0.1467 dB/km (2015) \cite{Makovejs2015} to finally 0.1419 dB/km (2017) \cite{Tamura2017,Tamura2018}. Initial losses in hollow-core photonic bandgap fibers (HC-PBGF) of 1.72 dB/km (2004) \cite{Mangan2004} and 1.2 dB/km (2005) \cite{Roberts2005} where significantly reduced over the last years by use of nested antiresonant nodeless fibers (NANF) down to 0.65 dB/km (2019) \cite{Bradley2019}, 0.28 dB/km (2020) \cite{Jasion2020} and even 0.174 dB/km (2022) \cite{Jasion2022}.  While for solid-core fibers the improvements slowly converges towards the theoretical minimum governed by Rayleigh scattering, HCFs and in particular antiresonant fibers show rapid development and might surpass solid core fibers in near future.}    
\end{figure}

In an optical fiber, precisely because the optical field lives in a medium, physical processes in the medium induce optical noises and nonlinearities which preclude the use of high powers to overcome the limitations posed by the already miniscule losses. It is in this context that structured optical fibers seem to offer a unique opportunity: structuring of the core offers a new degree of freedom which allows independent engineering of the optical noises and losses. In fact, just last year, a structured fiber with loss comparable to a solid-core has been demonstrated \cite{Jasion2022}. \Cref{Fig:Losses} depicts the dramatic progress in the reduction of optical loss achieved in structured fibers in recent years, while solid-core fiber technology seems to be limited by 40 years of incremental improvement. Thus, the question of whether optical fiber technology, given its mature state, can supplant  or supplement the conventional platforms employed in precision and quantum measurements needs a fresh appraisal. 

We first review the primary limitations of solid-core optical fibers as it pertains to applications in precision optical sensing, including the effect of optical noises and nonlinearities. We then discuss the state of knowledge of these same factors in structured optical fibers, and highlight opportunities for further research into their properties. Given the obvious practical advantage of optical fibers over free-space optical links --- immunity to electromagnetic interference, single-mode performance, and ease of deployment --- we expect that a careful study of the optical noise properties of structured fibers may renew interest in the application of optical fibers to precision quantum sensing and communication.
\section{Setup}\label{sec:setup}

To quantify the performance of large optical fiber interferometers, and ultimately be able to compare them to their free-space counterparts, a simple Mach-Zehnder interferometer (MZI) is considered (\cref{sec:setup}). We work out all known relevant sources of intrinsic noise in an optical fiber that can  affect the performance of such an interferometer. In particular, we review a panoply of noises intrinsic to optical fibers (\cref{sec:Intrinsic}) including scattering processes, thermodynamic noises, and noises due to nonlinear optical processes. We then briefly survey the major extrinsic sources of noise (\cref{sec:extrinsic}) that affect all fiber interferometers, including environmental perturbations and noise arising from the input optical field. The results we derive for this concrete example can be easily transferred to other topologies and applications that rely on phase-coherence.

\begin{figure*}[t!]
    \centering
    \includegraphics[width=0.8\textwidth]{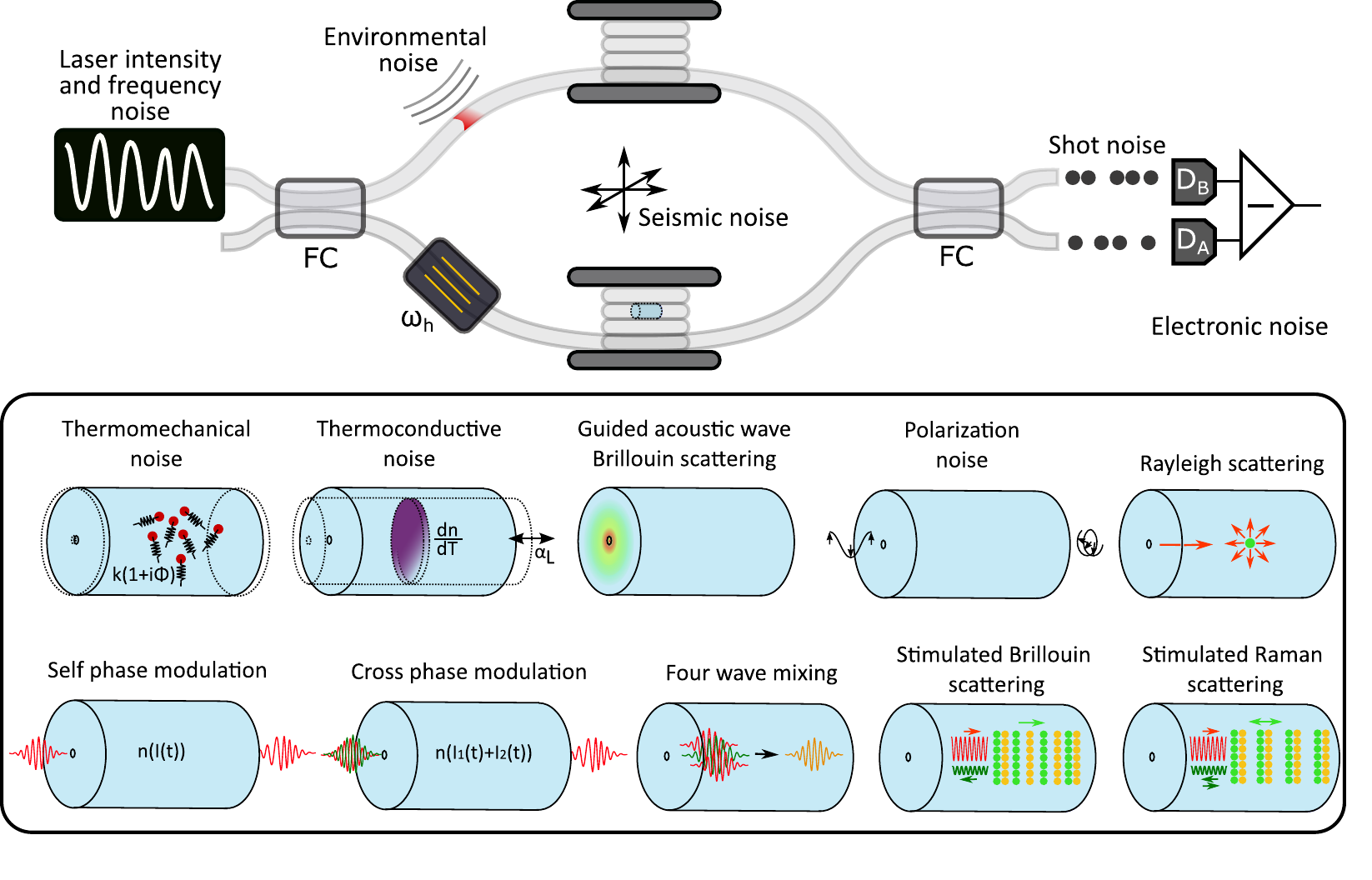}
    \caption{\label{Fig:Noises}
        Top panel shows a schematic fiber Mach-Zehnder interferometer probed by a noisy laser and
        perturbed by environmental noises (see \cref{sec:extrinsic}). 
        In addition to these extrinsic sources of noise that affect
        the interferometer, the optical fiber is itself a source of a panoply of noises, depicted schematically
        in the bottom panel (\cref{sec:Intrinsic}). The most important linear sources of intrinsic noise are shown in the first row and discussed in \cref{Sec: LinearNoise}, while we break down the nonlinear noise terms, shown in the second row, in \cref{Sec: NonLinear}. Thermodynamic phase noise, arising from a combination of mechanical- and thermal dissipation, is one of the main limitations in optical fiber interferometry and is especially important for long fibers at low ($\lesssim \SI{1}{kHz}$) 
        Fourier frequencies (\cref{Sec:Thermal}). 
        In addition, elastic scattering of light by transverse acoustic modes in the fiber, an effect known as guided acoustic wave Brillouin scattering (GAWBS), leads to a very complex phase noise spectrum with resonances occurring already in the low MHz range and a behavior at low frequencies that is not yet fully understood. Polarization noise in a fiber interferometer arises from birefringence fluctuations either in the fiber itself or the fiber connection to the laser (\cref{Sec:Polarization}). Rayleigh scattering, where light scatters off static density variations frozen into the fiber, becomes an important source of noise for narrow-linewidth lasers (\cref{Sec:Rayleigh}). For high input powers, nonlinear scattering effects, like self- and cross-phase modulation (SPM/XPM), add phase noise to the output signal, where the latter only occurs if multiple wavelength are multiplexed into the fiber. Four wave mixing (FWM) is also of particular importance for wavelength division multiplexing applications and introduces power losses depending on the phase matching between the involved light fields. Finally, the inelastic scattering of photons from acoustic (stimulated Brillouin scattering, SBS) or optical (stimulated Raman scattering, SRS) phonons restricts the transmittable power through an interferometer, where the former is the dominant source of nonlinear power loss (\cref{Sec: NonLinear}).
    }
\end{figure*}

We consider the Mach-Zehnder interferometer shown in \cref{Fig:Noises}: a fiber coupler (FC) splits the incoming light field --- prepared in a pure polarization state of the form $\vb{E}(t) = \vb{p}_0 E_0 \text{exp}(i \omega_0 t)$, with unit polarization $\vb{p}_0$, amplitude $E_0$, and angular frequency $\omega_0$ --- into two branches of lengths $L_1$ and $L_2$, respectively, before they are recombined at a second FC. This simple choice already discounts the spatial mode of the field, whose potential distortions in the fiber are not of interest to us. The outputs of the second fiber coupler are directed to a pair of photosensitive detectors whose output photocurrents are subtracted and recorded. To consider the possibility of heterodyne detection, one of the arms of the interferometer carries an ideal modulator, which shifts the frequency of light in that arm by $\omega_h$.

It can be shown that the optical power falling on the two detectors $A/B$ is

\begin{align}\label{eq:Pab}
    P_{A,B} (t) &= P_0\, R_{A,B} \left( 1 \pm V_{A,B} \cos \eta 
    \cos [\omega_h t + \phi(t)] \right),
\end{align}
where $P_0 \propto \abs{E_0}^2$ is the power sent into the interferometer. The splitting ratios of the two fiber couplers and the losses along each arm are captured by $R_{A,B}$, describing the transmissivity of each arm of the interferometer, and $V_{A,B}$, which is the ideal (perfect polarization overlap) interference visibility as measured by detector A/B. Note, that in general $R_A\neq R_B$ and $V_A\neq V_B$. The angle between the polarization states of light (in the Poincar\'e sphere) before they enter the combining fiber coupler is given by $2\eta$ , while $\phi(t) = \bar{\phi} + \delta \phi(t)$ is the phase difference between the two arms, consisting of a mean phase offset $\bar{\phi}$ --- possibly due to a length mismatch between the two arms --- and a fluctuating phase $\delta \phi$ --- consisting of the signal of interest, riding atop extraneous phase fluctuations from the interferometer. The difference photocurrent ultimately recorded is proportional to $P_A - P_B$. Thus, \cref{eq:Pab} suggests that noise in the photocurrent is primarily due to apparent fluctuations in the phase, polarization, and optical power in the interferometer.

The precise manner in which these fluctuations are transduced into the photocurrent depends on how the signal phase is extracted \cite{Giallorenzi1982,Sheem1982,Koo1982,Jackson1980-2,Dandridge1982,Cole1982,Nokes1978}. The simplest and most sensitive is active homodyne demodulation where the interferometer is maintained at quadrature (i.e. $\bar{\phi}=\pi/2$) using a feedback loop. For an interferometer with a large difference in the mean length between the two arms, this can be most easily achieved by controlling the laser frequency, while for a nearly balanced interferometer, by controlling the residual length difference. In either case, depending on the relative ratio of
the extraneous noise in the phase quadrature and environmental noise that drives the interferometer away from phase quadrature, the choice of gain of the feedback loop to stabilize the interferometer will dictate where the signal of interest is best extracted from. In the high-gain regime --- necessary when environment noise tends to destabilize the interferometer, and there is no separation in frequency between the signal and extraneous noise --- the signal is best extracted from inside the feedback loop. In the complementary low-gain regime, the signal can be extracted from outside the feedback loop \cite{Udd2011}. Heterodyne readout \cite{Dandridge1982} aims to produce a convenient frequency separation between signal and environmental noise so that the feedback loop can operate independent of the signal; the price to pay for convenience is sensitivity \cite{Yuen83,ShapWag84,BuoMav03}.

In the ideal case of (balanced) homodyne detection, all noise that obfuscates a phase signal of interest arises from various physical processes intrinsic to the optical fiber that produces apparent phase and polarization noise. In the following we tour the zoo of such noises. These include linear noise sources (in \cref{Sec: LinearNoise}) --- where the propagating optical field does not change the material properties of the fiber --- and non-linear noise sources (in \cref{Sec: NonLinear}). Finally, we discuss some noises extrinsic to the fiber (in \cref{sec:extrinsic}), which by dint of their insidious and ubiquitous character afflict all applications of precision fiber interferometry.

\section{Intrinsic noises in solid-core optical fiber interferometry}\label{sec:Intrinsic}

\subsection{Linear Noise Sources}
\label{Sec: LinearNoise}

\subsubsection{Rayleigh scattering}
\label{Sec:Rayleigh}

Perhaps the simplest mechanisms of optical noise in a waveguide is scattering from static surface roughness and density variation features, whose distribution in space is usually unknown. In an optical fiber, only a small fraction of the scattered light can be recaptured by it \cite{Nakazawa1983}. Rayleigh backscattered (RB) light then interferes with the main beam inducing phase noise in the output signal. The size of this noise largely depends on the coherence length of the light source \cite{Gysel1990}. The main concern for Michelson interferometers is the first order backscattering process, while Mach-Zehnder geometries suffer from double Rayleigh backscattered (DRB) light. For the MZI considered in this paper, the field at the end of each arm is the sum of the forward propagating main field and the double Rayleigh backscattered field. Assuming that the polarization does not change along the fiber (this represents the worst case scenario) and denoting the field at the first fiber coupler by $E(t)=E_0\text{exp}(i\omega_0t+i\delta\psi(t))\equiv E_0(t)\text{exp}(i\omega_0t)$, with $\delta\psi(t)$ denoting the laser phase noise (see \cref{sec:extrinsic}), the field reaching detector A can be written as 

\begin{align}
   E_\text{A}(t)=t_1r_2\Big[E_1(t)+E_1^\text{ds}(t)\Big]+r_1t_2\Big[E_2(t)+E_2^\text{ds}(t)\Big],
\end{align}
where $r_j$ and $t_j$ denote the complex amplitude reflection and transmission coefficients of the $j^\text{th}$-coupler. The forward propagating field amplitude at the end of the first (second) arm is given by $t_1 (r_1) E_i(t)$, with $E_i(t)=E_0\left(t-\tau_i\right)\text{exp}({-\frac{\alpha L_i}{2}})\text{exp}(i(\omega_0t-\beta L_i))$ and where $\tau_i=\frac{L_in_g}{c}$ denotes the group delay along arm $i$ with group refractive index $n_g$, $\alpha$ the (power) attenuation coefficient and $\beta$ the propagation constant. The calculation of the DRB field amplitudes $E_i^\text{ds}$ at the end of the $i^\text{th}$-arm can be done in a way similar to the RB fields \cite{Wan1996}. Backscattering can happen at any point $z_2$ along the fiber, while the location of double backscattering, $z_1$, is constraint by $z_1<z_2$. The DRB field thus has to cover an additional distance $2(z_2-z_1)$. The fraction of light scattered per unit length at random positions along the fiber is modeled by a complex white Gaussian zero-mean process, $\rho(z)$. The DRB fields can then be written as \cite{Wan1996}

\begin{align}
    E_i^\text{ds}(t)=\int_0^{L_i}&dz_2\int_0^{z_2}dz_1 E_0\left(t-\tau_i-\frac{2(z_2-z_1)n_g}{c}\right)\nonumber\\
    &\times e^{-(\frac{\alpha}{2}+i\beta)(L_i+2(z_2-z_1))}e^{i\omega_0 t}\rho(z_2)\rho(z_1).
\end{align}

The resulting noise is quantified by the autocorrelation of the optical intensity, $R_\text{DRB}=\langle I_A(t)I_A(t+\tau) \rangle$, where $I_A(t)=\epsilon_0 n c |E_A(t)|^2/2$ and $\langle\cdot\rangle$ denotes the ensemble average. The dominant noise contributions to $R_\text{DRB}$ are from the interference of the main fields with the DRB-fields in each arm ($E_i^\star E_i^\text{ds}$) as well as cross-interference terms between the main field of one arm with the DRB-field in the other ($E_i^\star E_j^\text{ds})$. The PSD of the former is the DRB-induced intensity noise of a single fiber. For light sources with a coherence length shorter than the fiber length it is given by \cite{Wan1996} (here and henceforth when we write a power spectral density it is single-sided)

\begin{align}
    S_\text{DRB}(\omega)=\frac{5}{9}\frac{\alpha_s^2S^2}{\alpha^2}\left(2\alpha L_i+e^{-2\alpha L_i}-1 
    \right)\mathcal{F}\{\langle |R_{E_i}(\tau)|^2\rangle\}\,,
\end{align}
where $\alpha_s$ is the Rayleigh scattering coefficient, $S$ the recapture factor, $\mathcal{F}$ the Fourier transform and $R_{E_i}$ the time autocorrelation of the forward propagating main field of the $i^\text{th}$-arm which depends on the coherence length of the light source. An extra factor of 5/9 takes birefringence effects in the fiber into account. Because DRB scatters a fraction of the fraction of forward propagating light, the effect of DRB on the output spectrum of a Mach-Zehnder interferometer is weaker when compared to the RB in Michelson interferometers. Note, that DRB in the input fiber lead couples laser frequency noise to intensity noise which adds to the relative intensity noise for narrow-linewidth lasers \cite{Fleyer2016,Cranch2003}.

\subsubsection{Thermodynamic phase noises}
\label{Sec:Thermal}

The fact that the optical fiber is in thermal equilibrium at a temperature $T$ produces fundamental thermodynamic fluctuations in its geometric and material properties that appear as apparent phase noise.

\subparagraph{Thermoconductive noise}\cite{Glen89,Wans92,Foster2007,Duan2012,Bartolo}. The dissipation of heat in the fiber material causes apparent fluctuations of the temperature. Since the phase acquired by light passing through a medium of (effective) refractive index $n$ and length $L$ is $k n L$ ($k=2\pi/\lambda$ is the magnitude of the wave vector), temperature fluctuations can manifest as apparent phase fluctuations via the temperature-dependent expansion of the fiber, and via the temperature-dependent changes in the refractive index. Thus the resulting phase noise is \cite{Foster2007},

\begin{equation}
\label{Eq:thermoCond}
    S_{\phi}^\t{TC}(\omega) = \frac{4 \pi^2 L^2}{\lambda^2}
    \left( \frac{dn}{dT} + n \alpha_L \right)^2 S_{\delta T}(\omega).
\end{equation}
Here $\alpha_L = (1/L)\dv*{L}{T}$ is the temperature coefficient of length expansion.

The apparent temperature noise $S_{\delta T}$ follows from the fluctuation dissipation theorem \cite{CalWel51,Kubo56} (FDT). The so-called ``direct form'' of the FDT \cite{Levin98} implies that any dissipated thermal power in the medium $W_\t{diss}$ results in a temperature fluctuation, $S_{\delta T} (\omega) = (8k_B T/\omega^2)(W_\t{diss}/Q_0^2)$, where $Q_0$ is the magnitude of the heat injected at frequency $\omega$. For a Gaussian optical beam with a mode field radius $r_\t{mf}$ propagating through a fiber core of thermal conductivity $\kappa$, the FDT implies \cite{Duan2012},

\begin{equation}
    S_{\delta T}(\omega) = \frac{k_B T^2}{\pi \kappa L} 
    \Re[e^{(i \omega r_\t{mf}^2)/4 D} \t{E_1}((i \omega r_\t{mf}^2)/4 D)].
\end{equation}
Here, $\t{E_1}(x) = \int_x^{\infty} t^{-1} e^{-t} \d{t}$ is the exponential integral function, and $D = \kappa/C_V$ is the thermal diffusivity (assuming a volumetric heat capacity at constant volume $C_V$). The above expression can be approximated by the so-called Wanser formula, \cite{Wans92,Foster2007}

\begin{equation}
\label{Eq:TempFluc}
    S_{\delta T}(\omega) = \frac{k_BT^2}{2 \pi \kappa L } 
    \ln{\left[\frac{k_\text{max}^4+\frac{\omega^2}{D^2}}{k_\text{min}^4+\frac{\omega^2}{D^2}}\right]},
\end{equation}
with $k_{\t{max}} = 2/r_{\t{mf}}$ and $k_{\t{min}} \approx 2.405/a$ ($a$ being the fiber outer  radius). A comprehensive table of constants can be found in the SI. 

\subparagraph{Thermomechanical noise} The thermoconductive noises ultimately arise due to thermal dissipation --- and associated temperature fluctuations --- in the active region of the optical fiber. By contrast, thermomechanical noise is due to mechanical dissipation, causing Brownian motion of the lengthwise elastic continuum 
that constitutes the fiber medium \cite{DuanGen}.

The 1D elastic continuum of the fiber length can be described as a sum of normal mode harmonic oscillators, each at a frequency, $\omega_\ell = (\ell \pi/L)\sqrt{E/\rho}$ ($E$ is the elastic modulus and $\rho$ the mass density), modal mass $m_\ell = \rho A L/2$ ($A$ the transverse area), and loss angle $\phi_\ell (\omega)$ (which are experimentally determined). Knowing the admittance of each normal mode in terms of these parameters, the FDT predicts an apparent length fluctuation due to thermomechanical noise,

\begin{equation}\label{eq:SLtm}
    S_{L}^\t{TM}(\omega) = \frac{4k_B T}{\omega} \sum_\ell \frac{\omega_\ell^2 \phi_\ell(\omega)}{m_\ell[(\omega^2 - \omega_\ell^2)^2+\omega_\ell^4 \phi_\ell^2(\omega)]} \,.
\end{equation}
If the loss angle is assumed frequency independent and uniform for all modes, i.e., $\phi_\ell(\omega) = \phi_0$, the above expression reduces, in the low-frequency regime (i.e. $\omega \ll \omega_\ell$), to

\begin{equation}
    S_{L}^\t{TM}(\omega) \approx \frac{4 k_B T L \phi_0}{3 A E} \frac{1}{\omega}.
\end{equation}
These length fluctuations can be referred to an apparent phase noise,

\begin{equation}
	S_{\phi}^\t{TM}(\omega) = \left( \frac{2 \pi n}{\lambda} \right)^2 S_L^\t{TM}(\omega).
\end{equation}
The frequency scaling suggests that thermomechanical noise dominates over thermoconductive noise at low frequencies, and ultimately limiting the long-term phase stability of long optical fiber links. The latter conclusion, however, relies on the poorly understood low frequency behavior of the loss angle $\phi(\omega)$. 

The sum of thermomechanical and thermoconductive noise, $S_\phi^\t{therm}=S_{\phi}^\t{TM}(\omega)+S_{\phi}^\t{TC}(\omega)$, is in excellent agreement with experimental results for frequencies down to about \SI{0.2}{Hz} \cite{Bartolo,Dong2016}. However, the resonances predicted by the thermomechanical theory have not apparently been observed, which, together with the measured deviations at lower frequencies, indicates the need for future investigations.

\subparagraph{Guided Acoustic-Wave Brillouin Scattering (GAWBS).} First observed in the context of quantum squeezing in optical fibers \cite{Shelby1985PRL,Shelby1985}, GAWBS refers to the scattering of light into the propagation direction by \emph{transverse} acoustic modes of the optical fiber. Unlike conventional Brillouin scattering (see \cref{Sec: NonLinear}), GAWBS is independent of the optical power in the fiber.

The dominant acoustic modes of the fiber responsible for GAWBS are the transverse radial ``breathing'' modes, and radial ``torsional'' modes with a near-zero longitudinal wave vector. The former mediates random scattering in the propagation direction --- through the photo-elastic effect --- akin to pure phase noise, whereas the latter can additionally produce polarization noise due to birefringence fluctuations. Consequently, they are labeled depolarized modes, whereas the pure 
radial components are called polarized modes. As far as phase noise is concerned, the primary contribution is due to the polarized modes (more so at low frequencies). Therefore, here we restrict attention to the polarized modes.

The elastic equations in the cylindrically symmetric frame of the optical fiber dictates the behaviour of the polarized modes. Assuming free boundary conditions, the various normal mode frequencies are given by, $\omega_\ell = (v_L/a)y_\ell$, where $v_L$ is the longitudinal acoustic velocity, $a$ is the fiber radius and $y_\ell$ satisfies the characteristic equation, $(1-\alpha^2)J_0(y_\ell)-\alpha^2 J_2(y_\ell)=0$, with $J_\ell$ the Bessel function of order $\ell$; here $\alpha = v_T/v_L$, and $v_T$ is the transverse acoustic velocity. Each frequency mode corresponds to the elastic radial motion of the fiber, which can be described by an eigenfunction, $U_\ell (r,t) =C_\ell(t) J_1(y_\ell r/a)$, whose amplitude, $C_\ell$, has the rms value $\sqrt{\langle C_\ell^2 \rangle}=\sqrt{k_B T/(m_\ell \omega_\ell^2)}$ due to the equipartition principle \cite{Sudhir2018}. Here, $m_\ell=\pi a^2 L \rho \int_0^1J_1^2(y_\ell x)x \d{x}$, is the effective mass of the mode and $L$ the fiber length. 

The photo-elastic effect determines the influence of this radial strain on the propagating optical field. Denoting by $P_{ij}$ the components of the photo-elastic tensor, the refractive index fluctuations are given by \cite{Shelby1985}: 

\begin{align}
    \label{Eq: GAWBSn}
        \delta n(r,t)= \sum_\ell \frac{n_\t{eff}^3(P_{11}+P_{12})C_\ell(t)}{2}
        \Big(&\frac{2}{r}J_1\left(\frac{y_\ell r}{a}\right)\nonumber\\
        &-\frac{y_\ell}{a}J_2\left(\frac{y_\ell r}{a}\right)\Big) \,.
\end{align}
The refractive index seen by the light is the average of this expression over the optical mode profile of the fundamental LP$_{01}$-mode of the fiber core. In a single-mode fiber, where most of the optical field is confined to the core, we have that $y_\ell r/a\ll\pi$, implying that the first- and second order Bessel functions in \cref{Eq: GAWBSn} can be approximated by $y_\ell r/(2a)$ and $y_\ell^2r^2/(8a^2)$, respectively; then the refractive index change can be approximated by
$\delta n\approx n_\t{eff}^3(P_{11}+P_{12}) \sum_\ell C_\ell(t)y_\ell/(2a)$. Referred to a phase noise, $\delta \phi_\text{GAWBS}^\text{radial}= (k L) \delta n = (kL) n_\t{eff}^3(P_{11}+P_{12}) \sum_n C_\ell(t)y_\ell/(2a)$, from which it follows that, 

\begin{equation}
\label{Eq:GAWBS}
    S_\phi^\t{GAWBS}(\omega)=\left(\frac{k L n_\t{eff}^3 (P_{11}+P_{12})}{2a}\right)^2 
    \sum_\ell y_\ell^2 S_{C_\ell} (\omega).
\end{equation}
Treating the elastic normal modes as harmonic oscillators with loss angles $\phi_\ell(\omega)$, their spectrum, $S_{C_\ell}$, has the form of a single term in \cref{eq:SLtm}. Just as in the case of thermomechanical noise, the question of the behaviour of the loss angle at low frequencies seems to be poorly understood. For example, if loss is assumed to be due to internal dissipation \cite{Saulson1990} (and thus $\phi$ being independent of $\omega$ over a large range of frequencies), then the phase noise due to GAWBS would exhibit a $1/\omega$ behaviour at low frequencies, whose magnitude however seems inconsistent with 
some low frequency measurements \cite{Dong2016,Bartolo}; on the other hand, a ``velocity-damped'' model for mechanical loss ($\phi \propto \omega$) would predict 
frequency-independent noise due to GAWBS, however such a loss model is inconsistent with fused silica \cite{Gonzalez1995}.

The treatment of the radial-torsional modes can be obtained analogously, with a different characteristic equation leading to a more complex spectral distribution.
The dependence on the azimuthal angle of these modes produces a fluctuating  birefringence in the fiber leading to depolarization \cite{Poustie1992,Poustie1993}.  

\subsubsection{Polarization noise}
\label{Sec:Polarization}

An ideal straight segment of fiber is cylindrically symmetric about its length, implying that the polarization state of light propagating in it is two-fold degenerate, i.e. any orthogonal pair of polarization states propagate through it with the same propagation constant. Real-world non-idealities such as imperfections during the manufacturing  process or extrinsic factors like stresses (from bending or twisting), and environmental perturbations can lift this degeneracy, leading to birefringence. Fluctuations in the birefringence will cause the polarization to fluctuate.

The polarization evolution along a fiber can be conveniently described as the action of a $4\times4$ Müller matrix on an input polarization state, $\hat{s}_i$, represented as a Stokes vector in a Poincar\'e ball \cite{Johnson1981}.  Since most modern fiber components show negligible polarization-dependent loss, the Stokes vector is confined to its surface. The transformation of the polarization state can then be visualized as a simple rotation. Each polarization state corresponds to a single point on the sphere's surface, where all linear polarization states lie on the equator and left- and right circular states are located at the two poles.
Within this picture it is easy to see, that whenever the input polarization is aligned with one of the two eigenvectors of the Müller matrix of the fiber, polarization is preserved during transmission \cite{Kersey1988}. For the narrow-linewidth light sources we consider here, there exist additionally a pair of preferred linear polarization states that are mapped to (different) linear output states at the fiber end \cite{VanWiggeren1999}. This dictates another orthogonal polarization basis pair which need not coincide with the polarization eigenstates of the fiber. Finally, there exists a third orthogonal polarization pair associated with the polarization mode dispersion character of the fiber \cite{Gordon2000}. These so-called principle states of polarization are defined such that their sensitivity to small optical frequency changes in the fiber is zero to first order \cite{Poole1986}.

In an interferometer, where each path is in general independent, the same input state is transformed to a different output state. Since the interference depends on their overlap, random birefringence fluctuations in the two arms or the input fiber cause variations in the visibility factor, the low-frequency part of this effect being called polarization-induced fading (PIF) \cite{Kersey1988,Kersey1988-2} (see SI for more details). Additionally, these fluctuations can induce apparent output phase noise which can be minimized by maximizing the visibility \cite{Kersey1988,Kersey1990,Kirkendall2004}. Several techniques have been documented to evade PIF:  (a) use of polarization-maintaining (PM) fibers, although PM fibers are unsuitable for precision interferometry due to their higher loss and dependence on operating conditions \cite{Frigo1984};  (b) polarization diversity reception \cite{Frigo1984}, where three independently polarized receivers are used to ensure that perfect interference (at DC) happens in at least one of them; the tradeoff is a significant reduction of contrast due to losses; (c) active control of the birefringence of the output fiber lead \cite{Wanser1987}, or input polarization \cite{Kersey1988-3}; (d) use of Farady mirrors (in a Michelson topology) \cite{Martinelli1989}. Some combination of these techniques may enable recovery of visibility.

A far more insidious effect is thermodynamic fluctuations of the birefringence of a non-ideal fiber (or a PM fiber) \cite{KrySud21}, which can manifest as broadband phase \emph{and} amplitude noise. For an amorphous material such as fused silica that forms the core of a single mode fiber, this variety of thermodynamic polarization noise is sub-dominant to the other thermodynamic phase noises. 

\subsection{Nonlinear Noise Mechanisms}
\label{Sec: NonLinear}

Nonlinear optical activity of fused silica --- the major component of the core of single mode optical fiber --- can lead to optical noises stimulated by the field propagating through it. Because the silica molecule possesses inversion symmetry, all even order electric susceptibilities must vanish. Thus the leading order optical nonlinearity is mediated by the third-order electric susceptibility $\chi^{(3)}$. As we shall now see, it suffices to describe a majority of relevant nonlinear noise effects found in optical fibers \cite{Agrawal2013}.

\subparagraph{Kerr Effect.} The optical intensity of light propagating down a fiber can change its refractive index:

\begin{equation}
    \label{Eq: Kerr-Effect}
    n(I) = n_0 + \frac{3 \Re{\chi^{(3)}}}{4\epsilon_0cn_0^2}  I \equiv n_0 + n_2 I;
\end{equation}
here $n_0 = \sqrt{1+\Re{\chi^{(1)}}^2}$ is the (effective) linear refractive index, and $I = P_0/A_\t{eff}$ is the optical intensity due to the (average) power $P_0$ focused in a transverse area, $A_\t{eff} \approx \pi r_\t{mf}^2$.

If light of a single frequency is propagating down the fiber, the intensity-dependent refractive index leads to self-phase modulation (SPM). This is captured by the additional phase, $\phi_{\t{SPM}}=k n_2 L_\text{eff} I$, accumulated over an effective fiber length, $L_\text{eff}=(1-\text{exp}[-\alpha l])/\alpha$, where the linear absorption coefficient $\alpha$ is measured in $m^{-1}$. (The effective length is the length of a lossless fiber having the same nonlinear impact as a lossy fiber with exponentially decaying intensity along its length. For modern commercially available fibers with attenuation coefficients as low as $\alpha \approx 0.16\,\text{dB}\,\text{km}^{-1}$,
the effective length is $L_\t{eff} \lesssim 27\, \t{km}$.) The SPM mechanism can transduce laser power fluctuations, $\delta P$, into phase fluctuations,

\begin{equation}
    \delta\phi_{\t{SPM}} =\gamma L_\text{eff} \delta P,
\end{equation}
where we combined the fiber properties for a given wavelength in the nonlinear coefficient $\gamma=kn_2/A_\text{eff}$. 

In an interferometer, the coupling between intensity fluctuations of the light source and output phase noise crucially depends on the splitting ratio of the first fiber coupler and the two arm lengths of the interferometer. For an interferometer with a perfect $50/50$ splitting ratio of the first fiber coupler one obtains

\begin{equation}
\label{Eq: PSD_SPM}
    S_\phi^\text{SPM}(\omega)=\left(\frac{5 \gamma P_0}{12}\right)^2 \bigr\lvert L_\text{eff,1}-e^{-i\omega \tau}L_\text{eff,2}\bigr\rvert^2S_\text{RIN}(\omega)\,,
\end{equation}
where $\tau$ denotes the time delay between the arms, $L_\text{eff,i}$ is the effective length of the $i^\text{th}$-arm, and $S_\t{RIN}$ is the relative intensity noise (RIN) spectrum. An extra factor of $5/6$ was introduced to account for the random polarization evolution in a standard fiber \cite{Stolen1978} (this factor is absent for polarization maintaining fiber arms). If the two arms of the interferometer are equal in length, i.e. $\tau=0$ and $L_\t{eff,1}=L_\t{eff,2}$, the fluctuations at the output are correlated and the noise due to SPM vanishes. For a finite but small length difference, $L_\text{eff,1}\approx L_\text{eff,2}=L_\text{eff}$, the square of the absolute value in \cref{Eq: PSD_SPM} reduces to $4L_\text{eff}^2\sin^2(\omega\tau/2)$. Finally, when one arm is 
substantially longer, e.g. $L_2\ll L_1\approx L_\text{eff}$, the same term reduces to $L_\text{eff}^2$.

If light of multiple frequencies is present in the fiber, the intensity of one field can modulate the phase of another, an effect called cross-phase modulation (XPM) \cite{Agrawal2013}. XPM is effective only for those fields whose spatial overlap is significant, which explains its significance in the core of an optical fiber.
For a pair of fields (labelled with indices $i,j$) of identical polarization, it can be shown that the refractive index experienced by the $i^\t{th}$ field is modulated by the $j^\t{th}$ field's intensity $I_j$ according to \cite{Agrawal2013},

\begin{equation}
    \label{Eq: n_XPM}
        n_i(I_j) = n_{0,i} +n_{2,i} I_i + 2 n_{2,i} I_j \,,
\end{equation}
capturing both SPM and XPM. This implies that the phase of the $i^\t{th}$ field is modulated as,

\begin{equation}
    \delta\phi_{\t{XPM}}^i = 2\gamma_{i} L_\text{eff} \delta P_j,
\end{equation}
where $\gamma_{i} = k n_{2,i}/A_\t{eff}$. In an interferometer, where the first fiber coupler has $50 \%$ transmission, the induced phase noise in the $i^\text{th}$-beam is related to the intensity noise of the $j^\text{th}$-beam via 

\begin{equation}
    \label{Eq: PSD_XPM}
        S_{\phi,i}^\text{XPM}(\omega)=\left(\frac{5 \gamma P_{0j}}{6}\right)^2 \bigr\lvert L_\text{eff,1}-e^{-i\omega \tau}L_\text{eff,2}\bigr\rvert^2S_\text{RIN}^j(\omega)\,,
\end{equation}
where again complete polarization scrambling is accounted for by a factor of $5/6$. 

Quite generally, any third-order nonlinear interaction between three waves producing a fourth in the process is called four-wave mixing (FWM). This mixing is of particular importance for wavelength division multiplexing, where the frequencies are equally spaced and significant cross-talk between the channels can occur \cite{Tkach1995}. FWM does not contribute noise in optical interferometry, but rather produces nonlinear power losses depending on the phases of the involved fields. If only two input fields are present (at frequencies $\omega_{1,2}$), secondary waves at frequencies $\omega_3=2\omega_1-\omega_2=\omega_1-\Delta\omega$ and
$\omega_4=2\omega_2-\omega_1=\omega_2+\Delta\omega$, with $\Delta\omega=\omega_2-\omega_1$, are created. These frequencies are of particular interest because they appear close to the ones of the two input fields, $\omega_1$ and $\omega_2$, provided that $|\Delta\omega| \ll \omega_{1,2}$. The power transfer from the primary to the secondary waves can be expressed as \cite{Hill1978,Shibata1987,Inoue1992}

\begin{equation}\label{eq:fwm:P3}
   P_3 = \eta \left(\gamma L_\text{eff}\right)^2 P_1^2 P_2 e^{-\alpha L},
\end{equation}
with the scattering efficiency,

\begin{equation}
    \eta = \frac{\alpha^2}{\alpha^2+\Delta\beta^2}
    \left(1+\frac{4e^{-\alpha L}\sin^2\left(\Delta\beta L/2\right)}{\left(1-e^{-\alpha L}\right)^2}\right) \,,
\end{equation}
depending strongly on the phase matching between primary and secondary waves. The power scattered into the fourth beam, $P_4$, is obtained by interchanging $P_1$ and $P_2$ in \cref{eq:fwm:P3}.

The phase mismatch for $\omega_3$ and $\omega_4$ can be written as $\Delta \beta_3=2\beta(\omega_1)-\beta(\omega_2)-\beta(\omega_3)$ and $\Delta\beta_4=2\beta(\omega_2)-\beta(\omega_1)-\beta(\omega_4)$, respectively. Taylor expanding the propagation constants up to third order in $\omega$ around the central frequency, $\omega_0=(\omega_1+\omega_2)/2$, results in \cite{Shibata1987,Buck2004}

\begin{align}
    \Delta\beta_{3,4}&=-\frac{d^2\beta}{d\omega^2}\Bigr\rvert_{\omega_0}\Delta\omega^2 \pm \frac{1}{2}\frac{d^3\beta}{d\omega^3}\Bigr\rvert_{\omega_0}\Delta\omega^3\nonumber\\
    &=\frac{2\pi c\Delta\lambda^2}{\lambda_0^2}\left(D(\lambda_0)\pm \frac{\Delta\lambda}{2}\frac{dD}{d\lambda}\Bigr\rvert_{\lambda_0}\right)\,,
\end{align}
with the plus sign for $\Delta\beta_3$ and the minus sign for $\Delta\beta_4$. To obtain the wavelength dependent expression, the dispersion parameter $D(\lambda)=d/d\lambda(d\beta/d\omega)=(-2\pi c/\lambda^2)d^2\beta/d\omega^2$ and $\Delta \lambda=\lambda_1-\lambda_2$ were used. The phase mismatch should be as large as possible in order to avoid efficient FWM and thus high nonlinear losses in the primary waves.

\subparagraph{Stimulated Brillouin and Raman Scattering.} 
Inelastic scattering of optical photons by acoustic or optical phonons, which represent the mechanical vibrations of the fiber medium, is another important class of non-linear effects in optical fibers. This manifests as optical loss into the mechanical modes. When the mechanical mode is at acoustic frequencies, the effect is called Brillouin scattering, while if the phonons are at optical frequencies it is termed Raman scattering. Their effect is two-fold: they primarily limit the optical power that can be employed but can also transduce noise from the phonon field onto the optical field. The consequence of an optical power threshold is that by limiting the usable power, the attainable signal-to-shot-noise ratio (at large Fourier frequencies, where shot noise can dominate over extraneous noises), which scales as $1/\sqrt{P}$, is limited.

The optical threshold due to SBS arises as follows. The pump field beats with the spontaneously back-scattered acoustic Stokes waves \cite{Kobyakov2010}. The resulting interference pattern changes the material density via electro-striction, creating a forward propagating refractive index grating from which more pump photons are back-scattered. As soon as a certain threshold pump power is reached, the Stokes wave builds up rapidly and limits the transmittable power. A good approximation to estimate the Brillouin threshold in standard single mode step-index fibers is \cite{Smith1972} 

\begin{equation}
\label{Eq: Pthres_Brill}
    P_\text{th,B}=\frac{\kappa K A_\text{eff}}{g_B L_\text{eff}}\,,
\end{equation}
where $g_B$ is the Brillouin gain coefficient, which is the value of the Lorentzian-shaped Brillouin gain spectrum at the Brillouin frequency $\omega_B$. For standard single-mode silica fibers $\omega_B \approx 2\pi \cdot 11\,\text{GHz}$ \cite{Ruffin2005} with $g_B\approx1.68 \times 10^{-11}$ mW$^{-1}$ at 1550 nm. The numerical factor $\kappa \approx 19$ accounts for modern low-loss optical fibers \cite{Kung2005}. The mixing efficiency $K$ between the pump and the Stokes wave depends on the input polarization as well as the properties of the fiber. For a completely scrambled input polarization $K\approx 3/2$, while for fibers with negligible birefringence and linear input polarization $K=1$ \cite{Deventer1994}. Note that this simple formula does not account for other fiber geometries, and the effective area needs to be replaced by the acousto-optic effective area \cite{Ruffin2004,Kobyakov2005}.

Stimulated Raman scattering (SRS), on the other hand, involves photon scattering off of molecular vibrations of the constituents of the fiber, which oscillate at $\omega_R \gtrsim 2\pi\cdot 10\, \t{THz}$. In this process forward and backward scattering of both Stokes and anti-Stokes light are possible, where the latter is
nevertheless typically much weaker due to the limited number of excited optical phonons at thermal equilibrium ($k_B T/\hbar \omega_R \ll 1$). In the absence of a second input field, with a frequency difference close to the molecular resonance frequency, the stimulated process builds up from noise. The interference between the two fields then drives the molecular resonances, creating a positive feedback loop leading to the stimulated process. Under the assumption of a Lorentzian gain profile, the threshold power, defined as input pump power for which Stokes and pump power are equal at the fiber output, for forward SRS is given by \cite{Smith1972}

\begin{equation}
    P_\text{th,R}=\frac{16pA_\text{eff}}{g_RL_\text{eff}} \,,
\end{equation}
where $p$ ranges from 1 for overlapping and 2 for completely scrambled polarization between the pump and the Stokes wave. The numerical factor of 16 needs to be replaced by 20 for backward scattering, which is why it is rarely observed in fibers due to the almost exponential increase of Stokes power beyond the threshold. The Raman gain coefficient, $g_R$, denotes the maximal value of the Raman gain spectrum. For standard single mode fibers, the peak occurs at a frequency difference between pump and Stokes wave of about $13.2$ THz, where $g_R\approx 1.1 \times 10^{-13}$ mW$^{-1}$ at 1550 nm \cite{Stolen1984}. Even though the gain coefficient is about two orders of magnitude lower than the one for SBS, the short lifetime of the optical phonons makes the gain spectrum for the Raman interaction about six orders of magnitude wider ($\sim40$ THz). This is a very attractive feature exploited in Raman amplifiers, where signal frequencies can be amplified over a broad range as long as the pump wave is chosen such that their frequency difference falls within the spectrum. 

From the two nonlinear scattering processes described above, SBS is in optical fiber interferometry by far more critical due to the larger gain coefficient, $P_\text{th,B}/P_\text{th,R}\approx g_R/g_B\approx 10^{-2}$. Beyond the threshold power, the backreflected Stokes wave experiences a very rapid growth while at the same time depleting the pump. Once a maximal level is reached, any additional incident power is reflected towards the front end of the fiber, limiting the usable power at the other end. Even though SBS does not directly contribute to phase noise, it affects other, power sensitive noise terms like shot noise. For long, high-performance optical fibers with $L_\text{eff}\approx\alpha^{-1}\approx27$ km, the Brillouin threshold for a pump wavelength at 1550 nm, $K=3/2$ and $\kappa=19$ predicted by Eq. \eqref{Eq: Pthres_Brill} is about 5 mW. Assuming the first beam splitter of the interferometer to have 50 percent transmission, the maximal laser power for long-baseline optical fiber interferometry is about 10 mW. 

The noise transduced by both effects are sub-dominant to other linear sources of thermodynamic noise. For Raman scattering this is due to the fact that the Raman-active phonons are not thermally populated at room temperature. For Brillouin-active phonons, their much larger frequencies, compared to the mechanical modes that participate in GAWBS etc, imply that their contribution is sub-dominant.

\section{Extrinsic Noises in Optical Fiber Interferometry}\label{sec:extrinsic}

\subsection{Source noise}

A real-world laser is a non-ideal oscillator whose output exhibits fluctuations in both its amplitude and phase. Noise in the amplitude is typical characterized through the fluctuations it causes in the optical power, normalized by the average output power of the light source, and is termed relative intensity noise (RIN). Such power fluctuations are converted into phase fluctuations and limit the sensitivity of optical interferometers; they are also known to limit the performance of optical communication systems \cite{AhmYam08}. In an interferometer, the transduction of RIN to output noise depends on the demodulation approach and arm length difference. Assuming the time delay difference between the arms to be much smaller than the correlation time of the power fluctuations, the RIN-equivalent phase noise for active homodyne detection at a single detector is given by $S_{\t{RIN}}/V^2$, which is at best as low as $S_\t{RIN}$ (when the visibility is $V=1$), and much worse in general. Balanced detection offers immediate mitigation of this stringent limit by subtraction of the large dc-component (see SI for details). Assuming 50/50 fiber couplers ($V_A=V_B\equiv V)$ and independent photodiodes with adjustable gains, $G_A$ and $G_B$, the RIN equivalent phase noise is given by 

\begin{equation}
    S_\phi^\t{RIN}(\omega)=\left(\frac{\epsilon}{2G_AR_AV}\right)^2 S_\t{RIN}(\omega)\,,
\end{equation}
where $\epsilon \equiv G_AR_A-G_BR_B$. Thus, the RIN-equivalent phase noise, proportional to the imbalance $\epsilon^2$, can be made arbitrarily small. This conclusion is crucially reliant on precisely maintaining the quadrature demodulation of the homodyne; for small deviations from this point will transduce RIN to apparent phase noise according to $S_\phi^\t{RIN}(\omega)=\tan^2(\Delta\phi) \, S_\t{RIN}(\omega)$, where $\Delta\phi$ is the phase deviation from quadrature, which is typically the dominant transduction process. RIN coupling can also be reduced by choosing a heterodyne detection scheme, where the dc-part is suppressed by the low-pass filter during the demodulation process, however at the expense of a 3 dB reduction in sensitivity.

Laser phase noise refers to phase fluctuations $\delta \psi(t)$ of the electric field,
$E(t) = E_0 \exp[i (\omega_0 t+ \delta \psi(t))]$. Assuming stationarity, it can be shown that
the electric field autocorrelation, $R_E(\tau)\equiv \avg{E^\ast(t)E(t+\tau)}$, 
is \cite{Elliott1982,DiDomenico2010},

\begin{equation}
    R_E(\tau)=E_0^2 
    \exp\left[i \omega_0 \tau -\int\limits_{-\infty}^\infty 
    S_{\delta \psi}(\omega) \sin^2 \left( \frac{\omega \tau}{2}\right) \frac{\d{\omega}}{2\pi} \right].
\end{equation}

It is typical to speak of ``frequency noise'', as the frequency-equivalent phase noise, defined by, $\delta \nu \equiv \dot{\delta \psi}/(2\pi)$, for which, $S_{\delta \nu}(\omega) = \left( \frac{\omega}{2 \pi} \right)^2 S_{\delta \psi}(\omega)$. A less informative, but common, measure of phase noise is the ``linewidth'', defined as the full-width at half maximum (FWHM) of the electric field power spectral density $S_{E}(\omega) = \int_{-\infty}^{\infty} R_E(\tau)\text{exp}(-i \omega \tau) \d{\tau}$.

Phase noise in the laser at the input of an interferometer can manifest as an apparent phase signal if there is an optical time delay $\tau$ between the two arms when they recombine. At recombination, the phase difference between the two fields due to laser phase noise is $\phi (t) = \delta \psi(t) - \delta \psi (t- \tau)$. Taking the Fourier transform gives the transfer function $(1-\text{exp}(-i \omega \tau))$ from $\delta \psi$ to $\phi$; consequently \cite{Arms66,GalDeb84,Kefelian2009},

\begin{align*}
 	S_{\phi}(\omega) 
 	&= 4 \sin^2 \left( \frac{\omega \tau}{2} \right) S_{\delta \psi}(\omega) \\
 	&= (2 \pi \tau)^2 \t{sinc}^2 \left( \frac{\omega \tau}{2} \right) S_{\delta \nu}(\omega).
\end{align*} 
where $\t{sinc}(x) = \sin(x)/(x)$. Thus laser phase or frequency noise directly manifests as an apparent phase signal $\phi$ sensed by the interferometer. At low Fourier frequencies, i.e. $\omega \ll \tau^{-1}$, $\text{sinc}(\omega\tau/2)\approx1$, so that, $S_{\phi} (\omega \ll \tau^{-1}) \approx (2 \pi \tau)^2 S_{\delta \nu}(\omega)$. The same relation holds for a heterodyne scheme, where the output oscillates at the modulation frequency before it is mixed down to dc in the phase-coherent detection. 

\subsection{Geophysical and environmental sources of noise}

Seismic ground motion is relevant at frequencies below $100\, \t{Hz}$. Its origins can range from the mundane --- such as anthropogenic and earthquakes --- to deeply fundamental aspects of the coupled dynamics of the earth's core, oceans, and atmosphere \cite{AkiRich09,Ard11,NakGuaFich19}. The $10-100\, \t{Hz}$ region is typically dominated by anthropogenic sources \cite{mcnamara_ambient_2004}, which exhibit significant diurnal and spatial variation, and is thus difficult to model. However,
because these disturbances are carried by surface waves, their propagation is attenuated over length scales of a few km. Earthquakes and their aftershocks dominate the ground motion at a few Hz. Below 1 Hz is a sequence of fundamental and universal spectra of seismic and atmospheric noises. The largest of these fundamental sources is the secondary microseism \cite{Hassel63,Ard11,Gual20} at $0.1-1\, \t{Hz}$ causing vertical ground motion of $\sim 10\, \mu\t{m}\cdot \t{Hz}^{-1/2}$; a factor of 5 smaller is the primary microseism at $\sim 0.02-0.1\, \t{Hz}$; another order of magnitude smaller is the seismic hum \cite{Webb07} at $\sim 1-20\, \t{mHz}$. Correlated with the microseism ground motion are microbarom atmospheric pressure fluctuations of $\sim (0.1-1)\, \t{Pa\cdot Hz^{-1/2}}$ at infrasonic frequencies \cite{Down67,Donn73}.

Ground motion causes strain on an optical fiber, which is transduced to apparent phase noise via the strain-optic effect. Assuming a fiber under isotropic stress, the phase change per unit fiber length and pressure ($p$) is approximately \cite{Hocker1979} $\delta\phi/(L \delta p)\approx \frac{\beta(1-2\nu)}{2E}\left(n^2(2P_{12}+P_{11})-2\right)$, where $E$ and $\nu$ denote Young's modulus and the Poisson ratio of the fiber, respectively, and $n$ the core refractive index. For modern fibers $\delta\phi/(L \delta p) \approx \SI{3e-5}{rad\, Pa^{-1} m^{-1}}$ at $\lambda=$ \SI{1.55}{\micro m}. If the fiber arms are spooled freely --- as a means to fold a long optical path length into a compact space --- and placed vertically, the spooled fiber experiences strain \cite{Ashby2007}, $\delta L/ L = (\nu_S \rho_S h/4E_S) \delta a$, where $\rho_S$ the density of the fiber spool, $h$ its height, $E_S$ the Young's modulus, $\nu_S$ its Poisson ratio and $\delta a$ the local vertical acceleration. The resulting phase noise is,

\begin{equation}
    S_\phi^\text{seis}(\omega) = 2\left(C_{\t{cmrr}}\frac{knL\nu_S\rho_S h}{4E_S}\right)^2 S_a(\omega),
\end{equation}
where $C_{\t{cmrr}}$ denotes the common mode rejection ratio accounting for the fraction of seismic noise coupling that is the same along both arms. If the fiber is spooled on an elastic core of length of several meters, atmospheric pressure fluctuations --- correlated across similar length scales --- can couple into the interferometer \cite{Zumb03}.

If the fiber interferometer is confined to a region of space smaller than the correlation length of some of these geophysical seismic noises, active vibration isolation can mitigate their effects \cite{Dong2016}. For applications where the majority of the optical fiber is outside of the laboratory, temperature changes along the fiber are typically among the most dominant sources of noise. Since both the effective refractive index and the fiber length are functions of temperature, the phase change per unit length and temperature change is \cite{Hocker1979} $\delta\phi/(L \delta T)\approx k[(n/L) dL/dT + dn/dT]$. Depending on the fiber used, even hundreds of radians per meter per degree are possible. In general, the combined phase noises of fibers exposed to urban environments over short time scales is Gaussian distributed \cite{Minar}.

Terrestrial gravity fluctuations \cite{Harms19} arise from the geophysical motion of large masses 
--- seismic density waves \cite{HughThor98}, atmospheric pressure waves, or even the motion of clouds \cite{Crei08} --- that couple
to the interferometer via direct Newtonian gravity. Because they are mediated by gravity, they cannot be shielded and thus represent the ultimate achievable limit of sensitivity on earth. 

\section{Intrinsic noises in structured-core fibers}

Due to technological advances and theoretical knowledge gained in the last few decades, optical fibers based on total internal reflection (TIR) have reached a point where they can only be marginally improved. The most important parameters for changing the properties in these structures are the choice of material and the concentration of added dopants. To overcome these limitations, optical fibers with alternative light guiding principles have been intensively researched over the last 30 years.

Photonic crystal fibers (PCF) are optical waveguides with a microstructured, periodic transverse refractive index profile \cite{Russell2003}. These microstructured fibers confine light to the core either by modified total internal reflection (M-TIR) \cite{Broeng1999}, photonic bandgaps (PBG) \cite{Yablonovitch1993,Birks1995,Cregan1999}, or by anti-resonant (AR) reflection \cite{Duguay1986}. Waveguides based on M-TIR, termed index guiding fibers, have a solid core surrounded by a periodic air-hole cladding with a lower average refractive index. By changing the size and separation between the holes they allow for much greater design flexibility as compared to TIR-based fibers \cite{Birks1997,Knight1998,Birks1999}. In hollow-core fibers (HCF), light is guided either via photonic bandgaps \cite{Yeh1978,Johnson2001,Poletti2013} or by the anti-resonance occuring at the glass membranes making up the boundary of the fiber core \cite{Couny2006,Wei2017}. Since light propagates mainly in air in these structures, low-loss guidance with reduced latency, higher damage thresholds and weaker nonlinear and thermal impacts is possible. 
 
The properties of photonic crystal fibers strongly depend on the transverse structure in addition to the choice of materials. Thus, they offer much more flexibility in their design and can be optimized for a wide range of different applications. In particular, HCFs appear to be an interesting alternative to conventional silica-based SMFs for optical fiber interferometry because more than 99 percent of the mode propagates through air \cite{Poletti2013}. 

\textbf{Rayleigh scattering and attenuation:} Due to the low overlap between the optical mode and the silica membranes of properly designed hollow-core photonic bandgap fibers (HC-PBGF), one can naively assume at least 20 dB reduction in the backscattering coefficient. However, mainly due to variations in the core dimensions and surface roughness induced by fundamental thermal noise, the actual scattering coefficient in those structures is higher than for conventional SMFs \cite{Roberts2005,Dangui2009,MichaudBelleau2021}. Since Rayleigh scattering is one of the main causes of loss at telecom-wavelengths, fibers with low attenuation coefficient are expected to show lower backscattering. Negative curvature fibers \cite{Wei2017} and in particular doubly nested antiresonant nodeless fibers (DNANF) \cite{Poletti2014} have nowadays attenuation coefficients as low as 0.174 dB/km \cite{Jasion2022} (S/C-band), and are expected to go below 0.1 dB/km in not too far future. This type of fiber achieved a backscattering coefficient of -118dB/km, which is more than 40 dB below the one for conventional solid core fibers \cite{MichaudBelleau2021} and is particularly useful to enhance the sensitivity of fiber optic gyroscopes \cite{Sanders2021}.

\textbf{Thermal noise:} Because the thermo-optic coefficient of air is much smaller than the one for fused silica, hollow-core fibers show a greatly reduced phase response to external temperature variations \cite{Slavik2019}. Indeed, the thermal sensitivity of HC-PBGF has been measured to be more than one order of magnitude lower than that of solid-core fibers \cite{Dangui2005,Slavik2015,Meiselman2017}. This is particularly important for applications where the fiber can't be shielded from the environment and the interferometric arms do not share common temperature drifts, e.g. for frequency dissemination. By cooling silica-based hollow-core fibers, it is even possible to build interferometers which are insensitive to external temperature drifts if operated close to the zero-crossing temperature of the thermal expansion coefficient of fused silica \cite{Zhu2019,Zhu2020}. This is not possible in conventional solid-core fibers, which are dominated by the thermo-optic effect and whose core and cladding regions have different doping concentrations with different thermal expansion coefficients. The thermal sensitivity can also be lowered by by reducing the thickness of the coating and thus its contribution to the thermal expansion, which has been shown to provide an almost 30-fold reduction of thermal sensitivity in DNANFs over solid-core fibers \cite{Shi2021}. Similarly, polarization-maintaining photonic crystal fibers show a larger birefringence and a more than 30-fold reduction in the temperature dependence of birefringence compared with PM solid-core fibers \cite{Ma2012}. In addition, the intrinsic thermodynamic fluctuations limiting the performance of conventional fiber interferometers are expected to be much smaller in hollow-core, air filled fibers and have been shown to already surpass conventional SMF interferometers for some frequency ranges \cite{Cranch2015}. 

\textbf{GAWBS:} Guided acoustic wave Brillouin scattering has been intensively studied in both solid-core PCFs \cite{Laude2005,Elser2006,Beugnot2007,Kang2009,Stiller2011,Jarschel2021} as well as HCFs \cite{Zhong2015,Renninger2016,Renninger2016_2,Iyer2020}. The GAWBS spectrum for the polarized and depolarized modes can be tailored by adjusting the transverse geometry of the fiber and thus the overlap between the optical and acoustic modes to either weaken its impact or enhance it for sensing applications. In HCFs, the spectrum contains contributions from the optomechanical coupling between the acoustic modes supported by the silica-capillaries (cladding) and air (core) with the optical modes \cite{Iyer2020}. Thus, the type of gas and its pressure are two additional degrees of freedom for tuning the optomechanical coupling. If low phase noise is required, evacuation of the core could be a viable solution. 

\textbf{Polarization:} Similar to conventional solid-core fibers, properly designed HC-PBGFs support two nearly degenerate, orthogonal fundamental modes \cite{Poletti2005}. Even though most of the optical power propagates through air, deformations in the transverse structure induced either during manufacturing or operation can lead to significant polarization mode splitting \cite{Bouwmans2003}. Intentional asymmetries in the structure surrounding the core can be used to produce polarization maintaining HC-PBGFs \cite{Roberts2005} with extinction ratios of more than 30 dB over hundreds of meters of fiber \cite{Fini2014}. With antiresonant fibers it is even possible to achieve extremely low polarization mode coupling without intentional birefringence, outperforming conventional and PM fibers by up to 3 orders of magnitude \cite{Taranta2020}. Additionally, this level of polarization purity is very robust against environmental perturbations, making antiresonant fibers an ideal candidate for high-precision fiber optic sensing.

\textbf{Nonlinear noise:} The ability to change the transverse geometry of photonic crystal fibers allows to engineer the size of the mode field and thus the intensity inside the fiber, which in turn governs the nonlinear behaviour \cite{Broderick1999}. HCFs also allow the strength of the nonlinearity to be modified by the filling material; when filled with air, the nonlinear interactions and associated phase perturbations are much smaller than those with an appropriate gas or liquid filling \cite{Dudley2009}. In particular, air-filled HC-PCFs have an effective nonlinearity that is three orders of magnitude lower compared to SMFs \cite{Bhagwat2008,Yang2020}. This allows to increase the threshold powers for SBS and SRS, which in turn enables higher signal powers to be delivered to the detectors. 

\section{Conclusion}

Solid-core optical fibers have come a long way since their introduction as transmission channels for communication systems in the 1970s. Thanks to continuous technical improvements, optical loss of contemporary solid-core single-mode fibers is only \SI{0.14}{dB/km}, which makes them very attractive for diverse applications. One particularly useful type is optical fiber interferometry, in which the fibers serve as both a transmission channel and as transducer for various external perturbations that change the properties of the propagating optical field. Depending on the application the transduction of these external parameters may manifest as signal or noise, and sensitivity to them may be desirable or extraneous. We have chosen to view the optical fiber as a conduit for phase-sensitive dissemination of light. This view encompasses applications ranging from distribution of precise frequency reference for optical clock networks, phase reference for quantum optical networks, and phase sensing for gravitational-wave detection. These applications are the most technically demanding deployments of optical phase stability
and sensing in the modern era.

\begin{figure}[t!]
\centering
\includegraphics[width=\columnwidth]{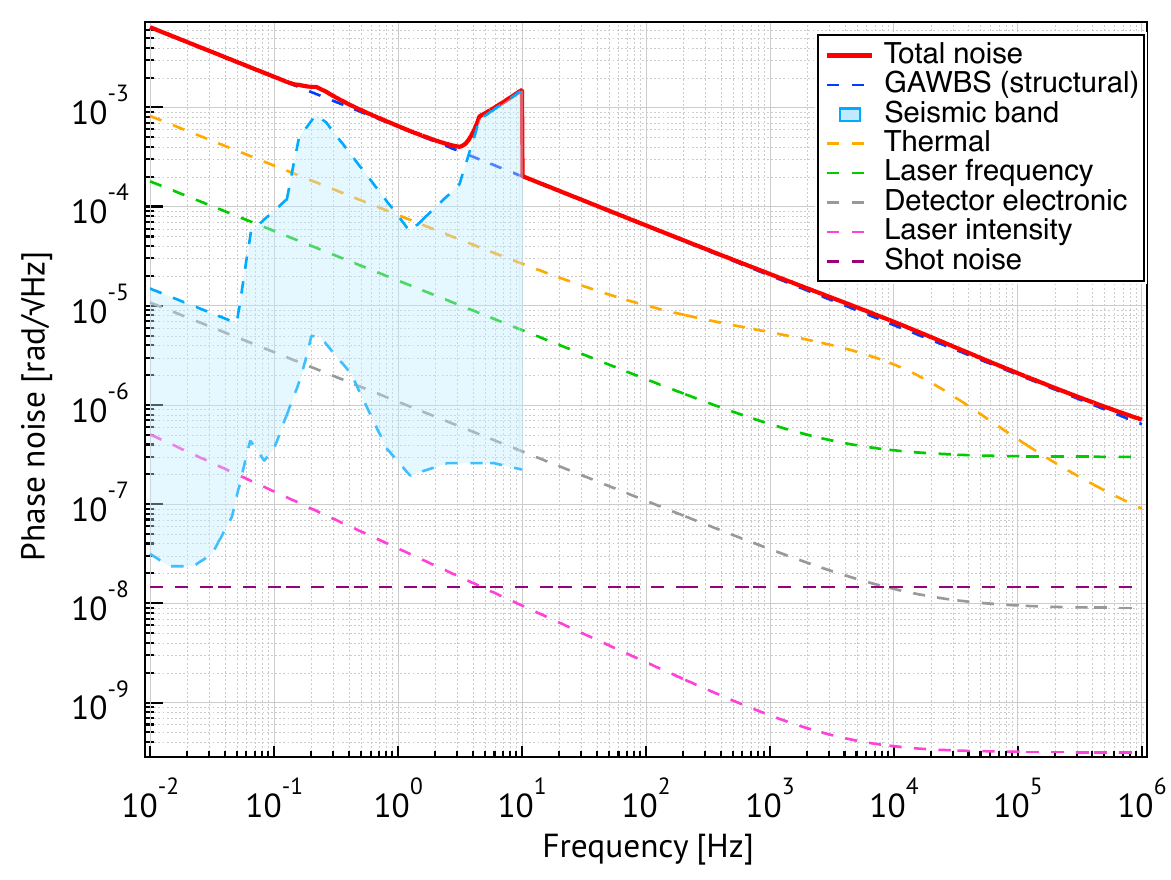}
\caption{\label{fig:budget}
	Projected low-frequency sensitivity of a fiber interferometer. 
	The phase stability of a \SI{10}{km} fiber interferometer, pumped with \SI{12}{mW} power
    (see text for details), with an arm length imbalance of \SI{1}{m}, 
    is primarily limited by a combination of thermodynamic and seismic noises at frequencies up to a MHz. 
	The primary thermodynamic contribution is from GAWBS, 
	here assumed to exhibit the typical structural damping mechanism.
	Other thermodynamic noises, originating from thermal dissipation,
	lurks less than an order of magnitude below. 
	Seismic noise is estimated using known models for the PSD distribution of global ambient
	seismic surveys \cite{peterson_observations_1993,mcnamara_ambient_2004,berger_ambient_2004}. 
    (Other relevant parameters used to produce this plot are available in the SI.)
}
\end{figure}

Under laboratory conditions (i.e. thermally and acoustically shielded), the sensitivity of an optical fiber interferometer at low frequencies is mainly limited by fundamental thermodynamic fluctuations and seismic noise. We illustrate the current typically achievable phase noise performance of an optical fiber Mach-Zehnder interferometer under ideal laboratory conditions in \cref{fig:budget}. With a fiber attenuation coefficient of 0.16 dB/km (best commercially available fiber), the maximal power per arm is restricted to \SI{12}{mW} right under the SBS threshold. That power also allows us to ignore nonlinear losses from SRS.
As the noise budget suggests, seismic and fundamental thermal phase noises dominate at frequencies below about\SI{100}{kHz}. Note however, that according to current theory, GAWBS appears to be the dominant source of noise, which, however, does not reflect the results obtained in thermal noise measurements, indicating a discrepancy in the understanding of acoustic losses that dictate GAWBS at low frequencies. (Indeed the ubiquitously-observed frequency-independent loss angle \cite{Saulson1990} assumed here is also inconsistent with causality.) Careful investigation of the broadband acoustic loss of the GAWBS mode is necessary to resolve this
inconsistency. A similar inconsistency exists for thermomechanical noise, whose predicted resonance peaks --- originating in the normal mode expansion (each mode is treated as harmonic oscillator) --- were not observed \cite{Bartolo,Dong2016}. In this plot, DRB is negligible due to the small recapture- and Rayleigh scattering coefficients. Assuming a short input fiber lead, the conversion between laser frequency to intensity noise due to DRB at the input can also be safely ignored. The relative intensity noise coupling is shown for a typical commercially available narrow-bandwidth fiber laser and assuming a balanced homodyne demodulation scheme, where the gains of the photodiodes are equal to within $\sim \SI{10}{ppm}$. For the arm length imbalance of $\Delta L=$\SI{1}{m} assumed here, the transduction of the frequency noise of a commercial fiber laser onto output phase noise is still sub-dominant to fundamental thermal noises. This however, is not true for significantly larger difference between the arms, 
e.g. for the largely imbalanced interferometer used for the dissemination of optical frequencies. For this type of interferometer, even the residual frequency noise of state-of-the-art optical local oscillators can dominate the phase noise at the output. Finally, at the relatively low input powers we assume here, together with the monochromaticity of the input light, and nearly equal arms lengths, the effects of SPM, XPM and FWM are also negligible.

One of the most demanding applications for such an interferometer (in a Michelson configuration) is gravitational wave detection. 
(In fact the optical fiber technology available in the 1980s was examined in this context \cite{Bluebook}.) Although a fiber-based interferometer would reduce the operational costs and provide other advantages --- like single spatial mode propagation, much reduced vacuum requirements, and compact geometries --- the obtainable phase sensitivity is still orders of magnitude worse than contemporaneous free-space multi-km interferometers (which realize \SI{1e-9}{rad/\sqrt{Hz}} at \SI{100}{Hz}). Even in the absence of thermal noise, SBS would limit the amount of usable power and thus establish an unacceptably high level of shot noise. Optical fiber interferometers can also potentially be used for other fundamental applications, like testing the interplay between general
relativity and quantum physics with light \cite{Tanaka1983,Zych2012,Rideout2012,Hilweg2017,Pallister2017,Rivera2021,Esmaeilifar2022}. Again, fundamental thermal noise sets the sensitivity limit which is, nevertheless, much less demanding compared to gravitational wave detection.

Long optical fibers are most frequently used outside the laboratory, for sending light signals between spatially separated locations. The noise caused by environmental perturbations typically exceeds thermodynamic noise by orders of magnitude, which is why in many practical applications the fiber is embedded in an interferometer for active phase stabilization. This allows, for example, the dissemination of optical frequency standards over distances up to almost 2000 km \cite{Droste2013,Droste2015,Raupach2015,Riehle2017,Hu2020}. A related field is relativistic geodesy, where the clock rates of optical frequency standards are compared via optical fiber links to gain information about differences in gravitational potentials between sites \cite{Grotti2018,Schioppo2022}. It is truly remarkable that actively stabilized optical fiber interferometers provide enough stability to use general relativistic time dilation as a tool to measure differences in altitude. Interferometric phase stability is also required for many quantum communication protocols \cite{Gisin2007,Minar}; for example in the recently much acclaimed field of twin-field quantum key distribution \cite{Chen2021,Wang2022}, which avoids the need for quantum repeaters between the sites.

Taken together, it seems that solid-core optical fiber technology is quite mature, with little room for further improvement in noise performance, except for a lack of complete understanding of thermo-optic losses at infrasonic ($< \SI{10}{Hz}$) frequencies. 

Rapid development in the past few years in the production of structured-core optical fibers with losses approaching those of solid-core fibers offers a potential roadmap for optical fiber based platforms for quantum technology and precision measurements. The scanty state of knowledge of the noise processes in these fibers is an open invitation for academic research. Although it is hard to imagine an orders of magnitude improvement in noise properties in such fibers, any improvement will break the barrier of diminishing returns observed in the solid-core fiber platform and will therefore produce new opportuninities for science and technology.

\section*{Acknowledgements}

We thank Robert Peterson, Raffaele Silvestri, Gesine Grosche, 
Thomas Waterholter, Geoffrey A. Cranch, Victoria Xu, and Evan Hall for useful discussions. 
P.W. acknowledges that this research was funded in whole, or in part, by the Austrian Science 
Fund (FWF) [TAI483], [F7113], [FG5], and through the research platform TURIS.
For the purpose of open access, the author has applied a CC BY public copyright licence to 
any Author Accepted Manuscript version arising from this submission.


\bibliography{References}

\end{document}